\documentclass[10pt,twocolumn,letterpaper]{article}

\usepackage{iccv}  
\usepackage{pifont}
\usepackage{float}
\usepackage{multirow}
\usepackage[normalem]{ulem}
\useunder{\uline}{\ul}{}
\definecolor{iccvblue}{rgb}{0.21,0.49,0.74}
\usepackage[pagebackref,breaklinks,colorlinks,allcolors=iccvblue]{hyperref}

\def\confName{ICCV}
\def\confYear{2025}

\definecolor{iccvblue}{rgb}{0.21,0.49,0.74}
\usepackage[pagebackref,breaklinks,colorlinks,allcolors=iccvblue]{hyperref}
\title{LD-LAudio-V1: Video-to-Long-Form-Audio Generation Extension with Dual Lightweight Adapters}

\author{
Haomin Zhang\textsuperscript{1}\thanks{\;These authors contributed equally to this work.}\hspace{0.6em} \quad
Kristin Qi\textsuperscript{2}\footnotemark[1] \quad
Shuxin Yang\textsuperscript{3}\footnotemark[1] \quad
Zihao Chen\textsuperscript{1} \quad
Chaofan Ding\textsuperscript{1} \quad
Xinhan Di\textsuperscript{1} {\ding{41}}\\
\textsuperscript{1}Giant Network, China \\
\textsuperscript{2}Computer Science, University of Massachusetts Boston, USA \\
\textsuperscript{3}Trine Univeristy, USA \\
{\tt\small zhanghaomin@ztgame.com, yanankristin.qi001@umb.edu, syang25@my.trine.edu,} \\ {\tt\small chenzihao@ztgame.com, dingchaofan@ztgame.com, dixinhan@ztgame.com}
}
\begin{document}
\maketitle
\begin{abstract}
Generating high-quality and temporally synchronized audio from video content is essential for video editing and post-production tasks, enabling the creation of semantically aligned audio for silent videos. However, most existing approaches focus on short-form audio generation for video segments under 10 seconds or rely on noisy datasets for long-form video-to-audio zsynthesis. To address these limitations, we introduce LD-LAudio-V1, an extension of state-of-the-art video-to-audio models and it incorporates dual lightweight adapters to enable long-form audio generation. In addition, we release a clean and human-annotated video-to-audio dataset that contains pure sound effects without noise or artifacts. Our method significantly reduces splicing artifacts and temporal inconsistencies while maintaining computational efficiency. Compared to direct fine-tuning with short training videos, LD-LAudio-V1 achieves significant improvements across multiple metrics: $FD_{\text{passt}}$ 450.00 $\rightarrow$ 327.29 (+27.27\%), $FD_{\text{panns}}$ 34.88 $\rightarrow$ 22.68 (+34.98\%), $FD_{\text{vgg}}$ 3.75 $\rightarrow$ 1.28 (+65.87\%), $KL_{\text{panns}}$ 2.49 $\rightarrow$ 2.07 (+16.87\%), $KL_{\text{passt}}$ 1.78 $\rightarrow$ 1.53 (+14.04\%), $IS_{\text{panns}}$ 4.17 $\rightarrow$ 4.30 (+3.12\%), $IB_{\text{score}}$ 0.25 $\rightarrow$ 0.28 (+12.00\%), $Energy\Delta10\text{ms}$ 0.3013 $\rightarrow$ 0.1349 (+55.23\%), $Energy\Delta10\text{ms(vs.GT)}$ 0.0531 $\rightarrow$ 0.0288 (+45.76\%), and $Sem.\,Rel.$ 2.73 $\rightarrow$ 3.28 (+20.15\%). Our dataset aims to facilitate further research in long-form video-to-audio generation and is available at \url{https://github.com/deepreasonings/long-form-video2audio}.
\end{abstract}
  
\section{Introduction}
Video-to-audio (V2A) synthesis, commonly known as Foley sound generation, represents a fundamental challenge in multimedia content creation that aims to generate semantically and temporally aligned audio for silent videos \cite{zhang2024foleycrafter}. This task demands not only understanding visual semantics and their relationship with audio to produce contextually appropriate sounds, but also ensuring precise temporal synchronization, as humans are highly sensitive to audio-visual misalignments as subtle as 25 milliseconds \cite{Lee2019ObservingPA,Koepke2020SightTS}.
\begin{figure}[t]
    \centering
    \includegraphics[width=1.0\linewidth]{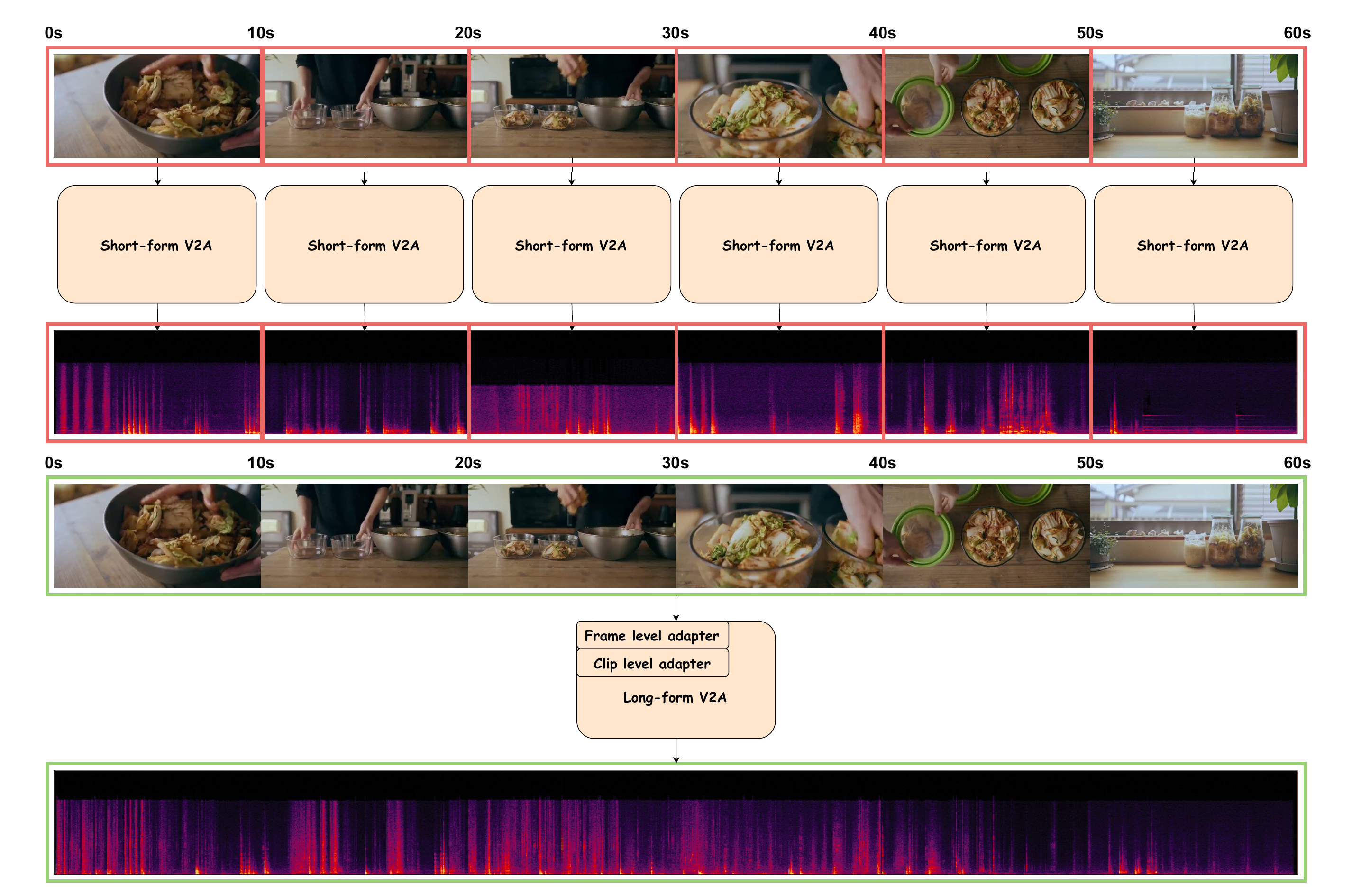}
    \caption{Long-form video-to-audio (V2A) generation with dual lightweight adapters.}
    \label{fig:head_longtime}
\end{figure}
\begin{figure*}[t]
    \centering
    \includegraphics[width=0.95\linewidth]{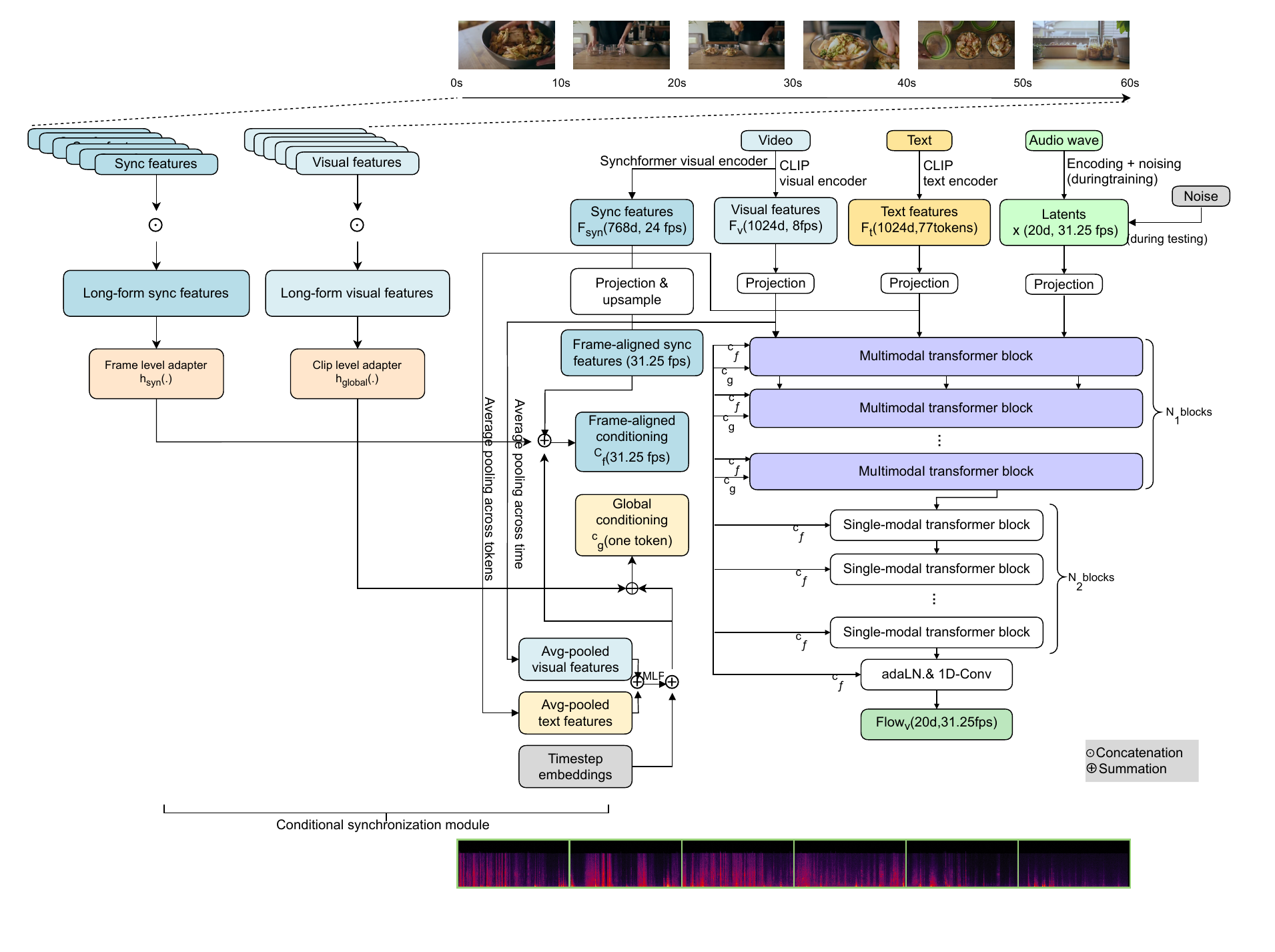}
    \caption{The framework details of LD-LAudio-V1 for long-form V2A generation.}
    \label{fig:v2a_longtime}
\end{figure*}
While recent advances in V2A synthesis have demonstrated capabilities in generating high-quality audio for short video segments \cite{luo2023difffoley,zhang2024foleycrafter,wang2024frieren,cheng2024taming}, existing approaches perform short-time audio generation, typically within $10$ seconds or less. Current V2A models have limitations when extending to longer temporal contexts. Commonly used approaches can be categorized into two main categories: autoregressive generation methods that produce audio in a sequential token-by-token manner~\cite{zhang2024foleycrafter,wang2023v2amapperlightweightsolutionvisiontoaudio,li2024tri}, and diffusion-based models with fixed-length denoising processes \cite{luo2023difffoley,wang2024frieren}. However, both categories contain challenges when adapted for long-time synthesis, particularly in maintaining accurate alignment between semantic and temporal domains when guided by visual information.

Recent research has attempted to address long-form audio generation challenges through various approaches, including diffusion transformer-based (DiT) architectures \cite{cheng2024lova} and multi-agent systems \cite{zhang2025longvideoaudiosynthesismultiagent}. Yet, these models have not been specifically designed to reduce splicing artifacts and temporal inconsistencies, nor do they demonstrate capability for high-quality generation across both short-form and long-form scenarios. Furthermore, high-quality long-form V2A datasets are lacking, with some being closed-source or containing voice and music noise \cite{geng2023dense}.

To address these limitations in long-form V2A generation, we first develop a high-quality V2A dataset containing pure audio effects that are human-annotated and free from voice and music noise. Second, we propose LD-LAudio-V1, a data-driven approach that extends short-form V2A synthesis models through dual lightweight adapters specifically designed to reduce temporal inconsistencies and splicing artifacts in long-form audio generation. As shown in Figure \ref{fig:head_longtime}, existing models segment videos, while ours uses dual lightweight adapters (frame- and clip-level).

\section{Related Work}
\subsection{Video-to-Audio}
V2A approaches can be categorized into autoregressive and diffusion-based categories. Autoregressive methods generate audio tokens sequentially~\cite{SpecVQGAN_Iashin_2021,viertola2024temporally,mei2024foleygen}, which are then decoded into audio signals. Latent diffusion and flow matching techniques have substantially enhanced Foley production quality and efficiency~\cite{luo2023difffoley,wang2024frieren}. Recent works of MultiFoley~\cite{chen2024video} combines mask denoising with reference audio for multi-modal control. MMAudio~\cite{cheng2024taming} utilizes a multi-modal transformer with flow matching and synchronization modules for enhanced temporal alignment. These studies inadequately address modality differences between audio and video and lack reasoning guidance. The newer work uses Factorized Contrastive Learning \cite{liang2024factorized} to enhance cross-domain alignment and propose a Chain-of-Thought (CoT)-like V2A approach, which facilitates both general V2A (VGGSound~\cite{chen2020vggsound}) and professional V2A (piano performance) through step-by-step guidance.

\subsection{Long-Form Video-to-Audio}
For long-form V2A synthesis, segmentation approaches like MMAudio divide extended videos into shorter clips for independent processing. These approaches contain temporal inconsistencies and loss of global context~\cite{viertola2024temporally}. Diffusion transformers such as LoVA \cite{cheng2024lova} demonstrate minute-level synthesis without explicit segmentation~\cite{chen2024video,geng2023dense}, and they demonstrate comparable performance in the comparison with the state-of-the-art models under heavy parameters trained with non-quantitation evaluation of inconsistency. A multi-agent based long-form V2A is developed with professional dubbing workflows through collaborative role specialization ~\cite{zhang2025longvideoaudiosynthesismultiagent}. Temporal fusion module emerges as a computationally efficient approach for long-form V2A generation ~\cite{ergasti2025rflav}. Similarly, Omni-based transformers are developed for end-to-end and fast V2A ~\cite{fei2025skyreelsaudio}. However, these long-form video-to-audio models do not demonstrate effectiveness in reducing inconsistency when generating long-form pure sound effects without noise. We curate a long-form V2A dataset and propose an extended model for generating multi-clip long-form V2A.

\section{Methods}
The long-form V2A task aims to generate an audio sequence $a$ of equivalent duration from a long video $v$. We propose LD-LAudio-V1, which extends the state-of-the-art models with dual lightweight adapters to handle multi-clip coherence. Our framework is shown in Figure \ref{fig:v2a_longtime}. It extracts global features from long-form multi-clip inputs and fuses them with short-form features to improve coherence. 

\subsection{Feature Representation}
We represent all features as one-dimensional tokens without using absolute position encoding, which allows generalization to different durations at test time. Visual features are extracted at $8$ fps as $1024$-dimensional features, and text features consist of $77$ tokens as $1024$-dimensional features, both extracted from CLIP \cite{radford2021learning}. Audio latents exist in variational autoencoder (VAE) latent space at $31.25$ fps as $20$-dimensional latents by default. Synchronization features are extracted with the Synchformer tool \cite{iashin2024synchformer} at $24$ fps as $768$-dimensional features. All features follow the same temporal ordering at different frame rates and are projected to a hidden dimension after initial processing layers.

\subsection{Frame-Level Synchronization Module}
For a sequence of video frames, we first extract per-frame visual features using a pre-trained vision encoder. Synchformer at $24$ fps as $768$-dimensional features. This synchronization module processes frame-level visual features to generate fine-grained temporal conditioning signals that are closely aligned with the temporal dynamics of audio. The Synchformer extracts features $F_{syn}$ and then performs projection and up-sampling to frame-aligned sync features $F_{syn}^{frame}$ with $31.25$ fps sampling rate.    


\begin{table*}
\centering
\resizebox{1.0\textwidth}{!}{%
\begin{tabular}{lccccccccccc}
\toprule
Method & $FD_{\text{passt}}\downarrow$ & $FD_{\text{panns}}\downarrow$ & $FD_{\text{vgg}}\downarrow$ & $KL_{\text{panns}}\downarrow$ & $KL_{\text{passt}}\downarrow$ & $IS_{\text{panns}}\uparrow$ & $IB_{\text{score}}\uparrow$ & DeSync$\downarrow$ & Energy$\Delta$10ms$\downarrow$ & Energy$\Delta$10ms(vs.GT)$\downarrow$ & Sem.Rel.$\uparrow$ \\
\midrule
GT & \textbackslash & \textbackslash & \textbackslash & \textbackslash & \textbackslash & \textbackslash & \textbackslash & \textbackslash & $0.1103$ & $0$ & \textbackslash \\
\midrule
MMAudio-L-44.1kHz Zeroshot & $455.49$ & $33.04$ & $2.28$ & $2.87$ & $2.32$ & $\mathbf{4.65}$ & $0.27$ & $1.44$ & $0.3629$ & $0.0524$ & $\mathbf{3.77}$ \\
MMAudio-L-44.1kHz Finetuned & $450.00$ & $34.88$ & $3.75$ & $2.49$ & $1.78$ & $4.17$ & $0.25$ & $\mathbf{1.38}$ & $0.3013$ & $0.0531$ & $2.73$ \\
MMAudio-L-44.1kHz Long-form & $\mathbf{\textcolor{LimeGreen}{327.29(-27.27\%)}}$ & $\mathbf{\textcolor{LimeGreen}{22.68(-34.98\%)}}$ & $\mathbf{\textcolor{LimeGreen}{1.28(-65.87\%)}}$ & $\mathbf{\textcolor{LimeGreen}{2.07(-16.87\%)}}$ & $\mathbf{\textcolor{LimeGreen}{1.53(-14.04\%)}}$ & $\textcolor{LimeGreen}{4.30(+3.12\%)}$ & $\mathbf{\textcolor{LimeGreen}{0.28(+12.00\%)}}$ & $\textcolor{OrangeRed}{1.51(+9.42\%)}$ & $\mathbf{\textcolor{LimeGreen}{0.1349(-55.23\%)}}$ & $\mathbf{\textcolor{LimeGreen}{0.0288(-45.76\%)}}$ & $\textcolor{LimeGreen}{3.28(+20.15\%)}$ \\
\bottomrule
\end{tabular}}
\caption{Results of long-form V2A generation with dual adapters compared to baselines.}
\label{tab:mmaudio_longtime}
\end{table*}

\subsection{Clip-Level Contextualization Module}
We integrate a clip-level contextualization module to produce a global semantic representation of the video. Features from the clip visual encoder $F_{v}$ and clip text encoder $F_{t}$ are first projected, then averaged and concatenated as $F_{g}^{con}$. The $F_{g}^{con}$ is fused with timestamp embeddings to produce global conditioning features $c_{g}$ that capture the semantic context of the video content.     

\subsection{Dual Lightweight Adapters for Coherence}
We extend short-form V2A to long-form audio generation through a multi-clip coherence extension module. This module applies the dual lightweight adapters, denoted as $h_{syn}$, $h_{global}$, to capture long-form consistency of audio from videos and then fuse them with the short-form videos.

\begin{itemize}
    \item \textbf{Light-weight Dual Adapters.} The long-form input multi-clip sequence from a video: $\{V_{clip}^{(i)}\}_{i=1}^L$ and is processed as a union long-form video from the same clip visual encoder and text encoder to extract the global visual features $F_{v}^{global}$, global text features $F_{t}^{global}$, and global synchronization features $F_{syn}^{global}$. These global features are processed similarly to $F_{v}$, $F_{t}$, and $F_{syn}$ to obtain global-level (multi-clip) conditions $c_{g}^{global}$ and local-level conditions $c_{f}^{global}$ after projection and average pooling (See Figure \ref{fig:v2a_longtime}).
    \item \textbf{Fusion.} For inference of audio for each video clip $V_{i}$, the final conditions are computed by combining global and local features through the lightweight adapters: $c_{g}^{final} = c_{g} + h_{global}(F_{v}^{global})$, $c_{f}^{final} = c_{f} + h_{syn}(F_{syn}^{global})$. 
\end{itemize}


\subsection{A Unified Multi-Modal Synthesis Transformer}
The same multi-modal transformer architecture used for short-form generation is applied to generate audio using the combined final conditions: $c_{g}^{final}$ and $c_{f}^{final}$. 

\begin{equation}
x^{(l+1)} = \text{DiT}^{(l)}\left(x^{(l)},\ c_{g}^{final},\ c_{f}^{final}\right),\quad \text{for } l = 1, 2, \dots, L
\end{equation}
where $x^{l}$ is the input of the $l_{th}$ transformer layers.

\section{Experiments}
\begin{figure*}[t]
    \centering
    \includegraphics[width=0.95\linewidth]{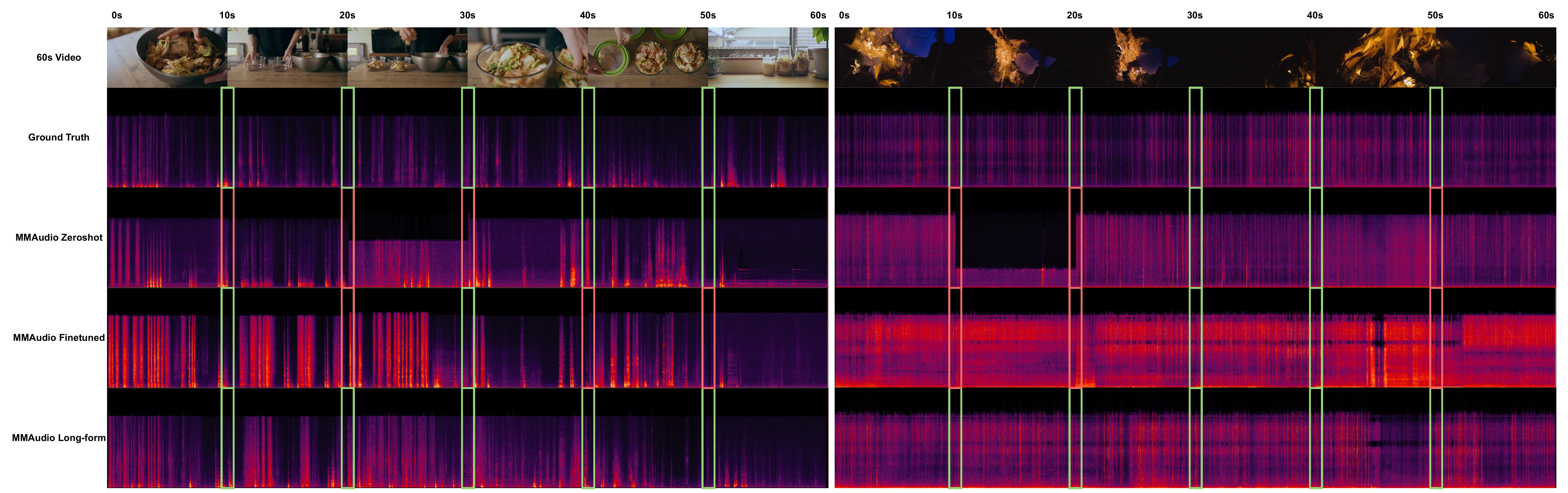}
    \caption{Examples of long-form V2A generation with different experimental settings.}
    \label{fig:specs_longtime}
\end{figure*}
\subsection{Datasets}
We curate the first version of a long-form clean sound effects dataset, denoted as LPSE-1, to support the study of long-form audio generation from videos. Different from the previous audio-visual event datasets (AVE) dataset, our LPSE-1 consists of more than $6K$ videos with over $20K$ audio-visual events covering $120$ different event categories. Each clip is more than $60$ seconds, containing real-life audio-visual scenes. Unlike other AVE datasets that contain noise such as voice-over \cite{chen2020vggsound} or other audio types such as music or speech, our long-form dataset contains pure sound effects. Each sample is manually human verified to ensure it contains only sound effects without other types of audio or irrelevant events from visual scenes.

\subsection{Evaluation Metrics}
\subsubsection{Quality Metrics}
Metrics for evaluating the quality of V2A are in four aspects, such as distribution matching, audio quality, semantic alignment, and temporal alignment \cite{zhang2024foleycrafter}. Specific metrics inlucde: $FD_{passt}$$\downarrow$, $FD_{panns}$$\downarrow$, $FD_{vgg}$$\downarrow$, $KL_{panns}$$\downarrow$, $KL_{passt}$$\downarrow$, $IS_{panns}$$\uparrow$, $IB_{score}$$\uparrow$, $DeSync$$\downarrow$ \cite{cheng2024taming}.

\subsubsection{Multi-clip Consistency Metrics}
Additionally, we apply consistency metrics specifically for long-form V2A. These metrics include the average energy change within 10 ms before and after each segmentation point between two short-form video clips ($Energy\Delta10ms$$\downarrow$), differences between the average energy change of the generated audio and ground truth ($Energy\Delta10ms(vs.GT)$$\downarrow$), and Semantic Relevance ($Sim.Rel.$$\uparrow$) \cite{cheng2024lova}.
\subsection{Initial Benchmark Results}
We compare our long-form model against the zero-shot MMAudio-L-44.1kHz model and a model finetuned on short training videos. The results are presented in Table \ref{tab:mmaudio_longtime}, with overhead costs in Table \ref{tab:info_longtime}. 

Our long-form model demonstrates significant improvements across multiple metrics compared to the short training videos finetuned model. Specifically, $FD_{passt}$ 450.00 to 327.29(+27.27\%), $FD_{panns}$ 34.88 to 22.68(+34.98\%), $FD_{vgg}$ 3.75 to 1.28(+65.87\%), $KL_{panns}$ 2.49 to 2.07(+16.87\%), $KL_{passt}$ 1.78 to 1.53(+14.04\%), $IS_{panns}$ 4.17 to 4.30(+3.12\%), $IB_{score}$ 0.25 to 0.28(+12.00\%), $Energy\Delta10ms$ 0.3013 to 0.1349(+55.23\%), $Energy\Delta10ms(vs.GT)$ 0.0531 to 0.0288(+45.76\%), and $Sem.Rel$ 2.73 to 3.28(+20.15\%). In Figure \ref{fig:specs_longtime}, we present several generated examples from different experimental settings.
\begin{table}[H]
\centering
\resizebox{1.0\columnwidth}{!}{
\begin{tabular}{lcc}
\toprule
Method & Params & Inference time of clip(60s) \\
\midrule
MMAudio-L-44.1kHz & 1.03B & 61.27s \\
MMAudio-L-44.1kHz Finetuned & 1.03B & 61.27s \\
MMAudio-L-44.1kHz Long-form with dual adapters & 1.07B & 62.75s \\
\bottomrule
\end{tabular}}
\caption{Computational costs of parameters and inference time.}
\label{tab:info_longtime}
\end{table}

\section{Conclusion}
We introduce LPSE-1, a clean dataset of 6k 60-second video clips with 24k verified audio-visual events, containing only pure sound effects. We also propose LD-LAudio-V1, which extends short-form V2A models to long-form audio generation using dual lightweight adapters, reducing splicing artifacts and temporal inconsistencies with only a 4\% parameter increase.

{
    \small
    \bibliographystyle{ieeenat_fullname}
    \bibliography{main}
}

\end{document}